\documentclass[]{spie}  

\usepackage{amsmath,amsfonts,amssymb}
\usepackage{graphicx}
\usepackage[colorlinks=true, allcolors=blue]{hyperref}

\title{Laboratory demonstration of an all-fiber-based focal plane nulling interferometer}

\author[a]{Jordan Diaz}
\author[a]{Rebecca Jensen-Clem}
\author[a]{Daren Dillon}
\author[a]{Philip M. Hinz}
\author[a]{Matthew C. DeMartino}
\author[a]{Kevin Bundy}
\author[b]{Stephen Eikenberry}
\author[b]{Peter Delfyett}
\author[b]{Rodrigo Amezcua-Correa}
\affil[a]{Univ. of California, Santa Cruz (United States)}
\affil[b]{Univ. of Central Florida (United States)}

\authorinfo{Further author information: (Send correspondence to J.D.)\\J.D: E-mail: jdiaz71@ucsc.edu}

\begin{document} 
\maketitle

\begin{abstract}
Starlight suppression techniques for High-Contrast Imaging (HCI) are crucial to achieving the demanding contrast ratios and inner working angles required for detecting and characterizing exoplanets with a wide range of masses and separations. The advent of photonic technologies provides new opportunities to control the amplitude and phase characteristics of light, with the potential to enhance and control starlight suppression. Here, we present a focal plane optical-fiber-based nulling interferometer working with commercially available components for amplitude and phase modulation. The instrument implements single-mode fiber-coupled elements: a MEMS variable optical attenuator (VOA) matches the on-axis and off-axis starlight amplitude, while a piezoelectric-driven fiber stretcher modifies the optical path difference between the channels to achieve the $\pi$ phase shift condition for destructive interference. We show preliminary lab results using a narrowband light source working at 632 nm and discuss future opportunities for testing on-sky with the Astrophotonics Advancement Platform at Lick Observatory (APALO) at the Shane 3-m Telescope. 
\end{abstract}


\section{INTRODUCTION}
\label{sec:intro}  

As of now, over 5600 exoplanets have been confirmed by means of different detection techniques \cite{exo_archive}. Indirect techniques, such as the transit method, have proven to be the most prolific, leading to approximately 98\% of these detections. On the other hand, direct imaging has the capability to collect light directly from companions, enabling spectroscopic measurements of their atmosphere, which are crucial to revealing chemical composition, understanding planet diversity and formation, and providing potential evidence of life \cite{Currie2023}. However, direct imaging poses a significant challenge: the detection of an Earth-like planet close to its host star at visible wavelengths demands a $10^{-10}$ contrast ratio \cite{Currie2023, Traub2010}. Therefore, significant efforts to develop and implement High-Contrast Imaging (HCI) techniques have been made, generally constituted by (1) incorporating adaptive optics (AO) systems into large aperture telescopes in order to counteract wavefront aberrations induced by atmospheric turbulence and the telescope+instrument itself and increase the angular resolution, (2) starlight suppression elements capable of significantly reducing the point spread function of the star that would otherwise outshine their companions, and (3) complex image processing algorithms \cite{Macintosh2014}.

Although the combination of AO and large diameter telescopes enables reaching angular resolutions close to the diffraction limit $\sim\lambda/D$, starlight suppression is essential to enabling the capability of distinguishing potential companions from the light of their host stars. One way to tackle this problem has been through the use of physical coronagraphs,  which working in tandem with powerful extreme AO systems, have led to confirming the detection of multiple exoplanets \cite{Macintosh2014}. Despite the progress made in exploring various coronagraph concepts and designs, contrast levels reached so far are limited to $10^{-6}$ at the inner working angles (IWA), i.e. the closest separation at which the throughput reaches half its power, $>$$\lambda/D$ - enough to resolve young, hot giant gas planets orbiting far from their host stars, but far from those required for rocky planets located in the habitable zone of their stars \cite{Habex2020}.

A second approach to suppress starlight is by means of nulling interferometry. This technique typically combines the light from two (or more) separate telescope apertures or sub-apertures pointing at the same star, and by arranging the $\pi$ phase shift difference between the beams, these interfere destructively \cite{Bracewell1978, Serabyn2021}. As a result, the on-axis starlight is nulled, whereas the light of an off-axis nearby companion, not meeting the destructive interference condition, remains.	First proposed by Bracewell \cite{Bracewell1978} in 1978, there has been a growing interest in the development of nullers during the last two decades due to the advancement of technologies such as photonic beam combiners and a promising advantage over coronagraphs: they have no fixed IWA. The IWA of a nuller depends on the baseline B, that is, the distance between the apertures, as $IWA=\lambda/2B$; enabling angular separations smaller than the diffraction-limited coronagraphs $\sim\lambda/D$ \cite{Serabyn2021, Martinod2021}.

Remarkable progress in the field of photonics during the last four decades, particularly driven by telecommunications, has yielded new commercially available products that could benefit high-contrast imaging and nulling interferometry.
 The promising advantages of miniaturization, scalability, and cost reduction have drawn astronomers' attention and given rise to the field of ``astrophotonics." A prime driver to develop astrophotonic instruments is the inherent ``spatial filtering" properties of single-mode (SM) waveguides from which spectroscopic and interferometric instruments would drastically benefit \cite{Jovanovic2016}. While feeding light coming from a telescope into SM waveguides raises a technical challenge, promising developments and progress have been made during the last decade. The advancement of extreme AO systems on large telescope facilities delivers diffraction-limited point spread functions capable of achieving on-sky coupling efficiencies $>40\%$ in the NIR \cite{Jovanovic2017}. Also, the multi-mode (MM) to SM transforming capabilities of photonic lanterns provide another venue for implementing photonic devices \cite{Birks2015}. Moreover, SM transportation of light enables a wide range of photonic functionalities, such as spectral dispersion, spectral, spatial, and polarization filtering, phase and amplitude modulation, frequency shifting, and light detection and generation, offering an unprecedented potential to enhance the effectiveness of high-contrast imaging instruments such as nulling interferometers \cite{Gatkine2019a, Jovanovic2023}. 

In this paper, we present preliminary results of the proof-of-concept of an all-fiber-based focal plane nulling interferometer comprised of commercially available components. The goal of this paper is to explore the feasibility of implementing the concept of nulling interferometry, taking advantage of the ease of accessibility to relatively low-cost optical fiber components. A description of the instrument's design, the laboratory configuration, and the main components and devices is given in Sec. \ref{sec:2}. Results on the performance of the instrument and an analysis of the barriers limiting its performance are provided in Sec. \ref{sec:3}. Finally, we present our conclusions and discuss possible future directions in Sec. \ref{sec:4}.

\section{DESIGN, AND LABORATORY SETUP}
\label{sec:2}
\subsection{The nulling interferometer's design}
\label{subsec:2-1}
A schematic of the interferometer is shown in Fig. \ref{fig:setup}(a). The instrument's design is based on the architecture of an optical fiber Mach-Zehnder interferometer, which collects light from two locations in the focal plane and injects it into two separate single-mode fibers: the first beam coupling light on-axis from the host star and the second from a nearby companion also containing contaminating starlight. In order to ensure and optimize the destructive interference condition, an amplitude and a phase modulation component are introduced downstream. The optical fiber transmitting the on-axis starlight is coupled in series to an amplitude (intensity) modulator to reduce it to the lower levels of contaminating starlight transmitted in the companion's arm. Before the beams are combined, optical fiber couplers with splitting ratios of 90/10 deliver the 10\% of each beam to reference photodetectors (PDs) for monitoring of the intensities just prior to beam combination. Then, beam combination results in two output beams with a $\pi$ phase shift between them: the one meeting the destructive interference condition is monitored as the ``nulling fringes`` channel, while the second yields the ``bright fringes`` channel.
\subsection{Laboratory configuration}
\label{subsec:2-2}
\begin{figure} [ht]
   \begin{center}
   \includegraphics[width=17cm]{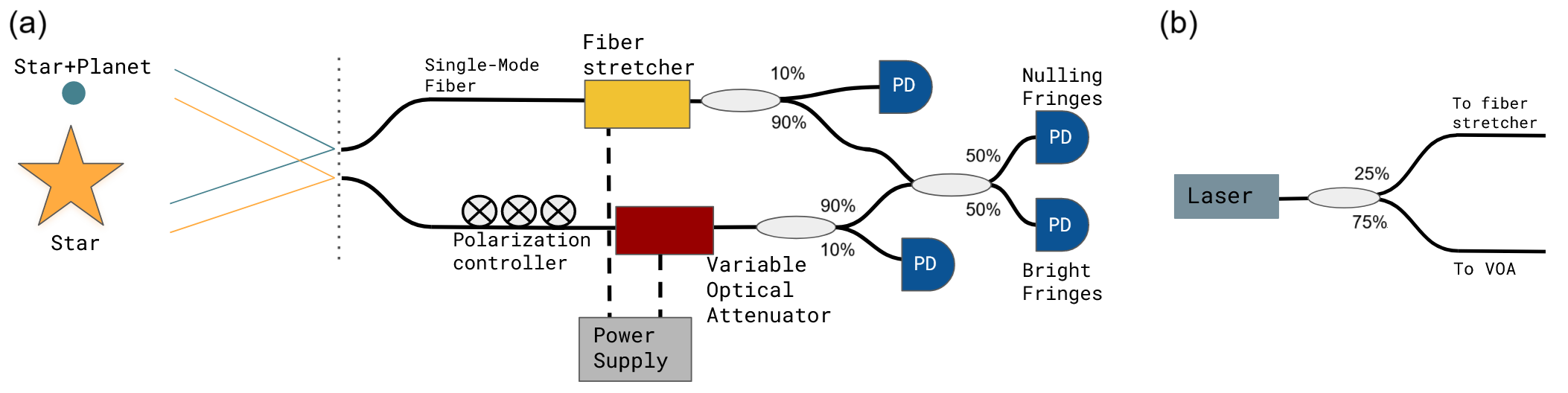}
   \end{center}
   \caption[example] 
   { \label{fig:setup} 
(a) Schematic of the all-fiber-based nulling interferometer and its components.(b) Light injection method implemented for the laboratory proof-of-concept. Monochromatic light from a laser is split using an optical fiber coupler with a 25/75 splitting ratio.}
\end{figure}

For our laboratory proof-of-concept, we simulated the two input channels with a monochromatic laser source and a 75/25 coupler, as shown in Fig. \ref{fig:setup}(b). We utilized a Thorlabs HNLS008L He-Ne laser with a central wavelength $\lambda$=632.8 nm and a linearly polarized output of $\sim$1mW. With the exception of the phase modulation unit, all optical fiber components used 630HP single-mode fiber optimized for the emission wavelength of the laser. To reduce laser fluctuations resulting from back-reflections from coupling light into the SMF, injection of light was done at a small angular deviation, which resulted in a coupling efficiency of $\sim$35\%. After the light was split, a manual three-paddle polarization controller was deployed in the star's arm for interference optimization. The components selected for amplitude and phase modulation were a variable optical attenuator (VOA) and a fiber stretcher, respectively, which are described in Section \ref{subsubsec:2-2-2}.

\subsubsection{Monitoring and control}
\label{subsubsec:2-2-1}
In the same vein as exclusively using off-the-shelf components for the interferometer, we also employ devices commonly found in a laboratory for monitoring the outputs and controlling the amplitude and phase modulation. The optical power of all four outputs was monitored using Thorlabs DET36A silicon photodetectors with a responsivity of 0.32 at $\lambda=632$ nm.  The photocurrents were visualized as voltages on a Tektronix MDO3024 oscilloscope. Communication with the oscilloscope was established with a PC  via Python through the \texttt{PyVISA} package, which serves as an interface for remote controlling of laboratory instruments \cite{Grecco2023}. A readout speed of $\sim$10 Hz for the four outputs (simultaneously) was achieved when the interferometer was not under active control, that is when no commands were sent to the power supply.

Both the VOA and the fiber stretcher were controlled using a GW Instek GPP-4323 programmable power supply. Communication with the power supply was also established through \texttt{PyVISA}. The power supply can deliver voltages in the 0-32 V range with a 1 mV resolution.
\subsubsection{Amplitude and phase modulation components}
\label{subsubsec:2-2-2}
\begin{figure}[h]
   \begin{center}
   \includegraphics[width=15cm]{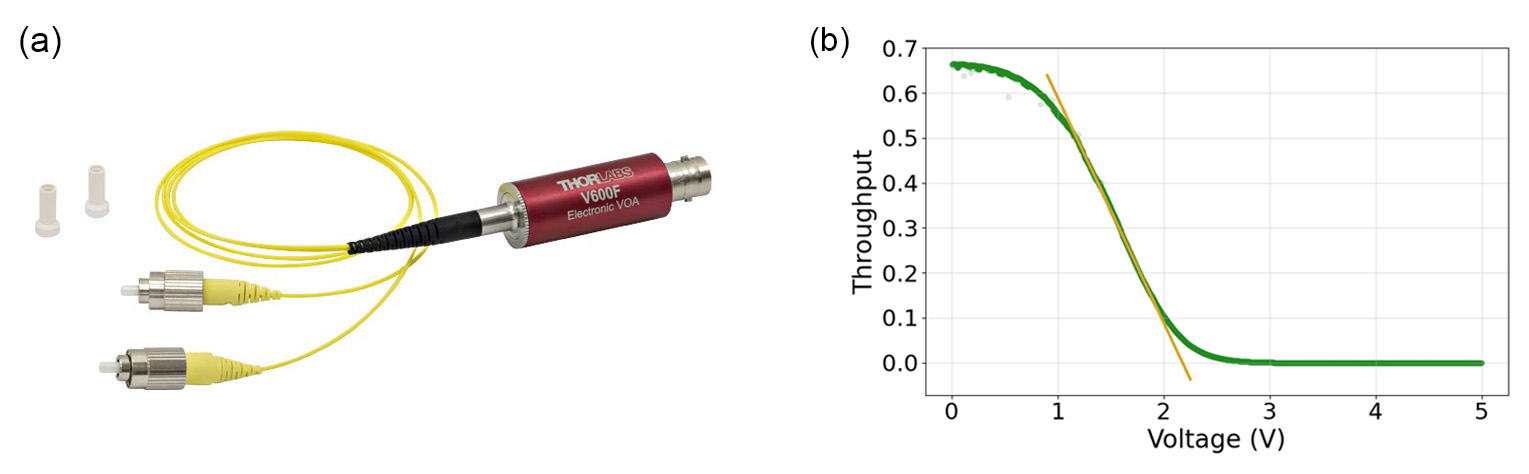}
   \end{center}
   \caption[example] 
   { \label{fig:voa1} 
(a) The Variable Optical Attenuator. Image taken from Thorlabs. (b) Throughput of the VOA as a function of input voltage.}
\end{figure}

The optical power transported in the star's arm of the interferometer was modulated using a Thorlabs V600F variable optical attenuator, shown in Fig. \ref{fig:voa1}(a). The VOA incorporates a micro-electro-mechanical system (MEMS) chip containing a movable tilting mirror that couples light between an input and an output fiber. The degree the mirror is tilted changes with the application of an input voltage, thus changing the coupling efficiency and the throughput of the device, allowing attenuation levels $>$2.5 dB up to 30 dB. 

The VOA's throughput as a function of input voltage is shown in Fig. \ref{fig:voa1}(b). Operation of the device is limited to voltages between 0-5 V, showing a nonlinear response with a linear window between $\approx$1-2 V. Note that the VOA shows an inherent coupling loss of $\approx$33\%. With the power supply's resolution of 1 mV, the device yields changes in throughput of 0.0075\% in the linear range of operation.

\begin{figure} [h!]
   \begin{center}
   \includegraphics[width=15cm]{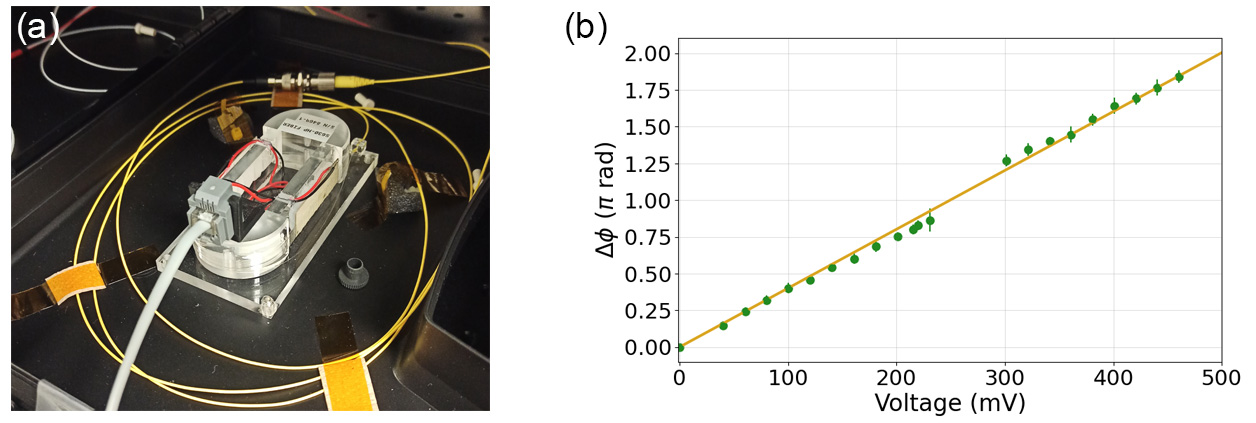}
   \end{center}
   \caption[example] 
   { \label{fig:fs1} 
(a) The piezo-electric fiber stretcher. (b) Characterization of the resulting phase shift $\Delta\phi$ as a function of input voltage. For each data point, several fixed phase shifts (same voltage) were applied when the interferometer output was at a minimum, from which a mean value and its standard deviation were computed.}
\end{figure}

Phase modulation was performed using a piezo-electric fiber stretcher from Evanescent Optics Inc. Model 915B. The device, shown in Fig. \ref{fig:fs1}(a), consists of an oval-shaped piece of plastic with two piezo-electric wafers mounted on its sides. Bare fiber is wound around the tube, and by feeding a voltage to the wafers, these can stretch or compress the fiber, consequently changing the optical path length and resulting in a relative phase shift between the arms of the interferometer. Our device has 19 wraps of S630-HP fiber with a maximum fiber stretch of 9 $\mu$m, equivalent to a maximum phase shift of $\sim$28 $\pi$. The fiber stretcher is not directly controlled by the power supply. A dedicated piezo-electric controller, Evanescent Optics Inc. Model 914-1, is implemented between the power supply and the fiber stretcher to drive the piezo-electric wafers. 

The phase shift $\Delta\phi$ resulting from the path length change was characterized as a function of input voltage - see Fig. \ref{fig:fs1}(b). The experiment to characterize it consisted of selecting a fixed voltage that was applied to the fiber stretcher when one of the monitored interference outputs was close to its minimum value. Several of these voltages were applied and then converted into a phase shift by estimating the change in the amplitude of the sinusoidal output intensity of the interferometer - $I_{out}\propto \cos(\phi)$. The process was repeated for different voltages, finding a linear relationship of $\Delta\phi=0.004\pi$ rad/mV, equivalent to path lengths of $\sim$1.2 nm.

\section{PERFORMANCE}
\label{sec:3}
\subsection{Response to environmental fluctuations}
\label{subsec:3-1}
\begin{figure} [h!]
   \begin{center}
   \includegraphics[width=17cm]{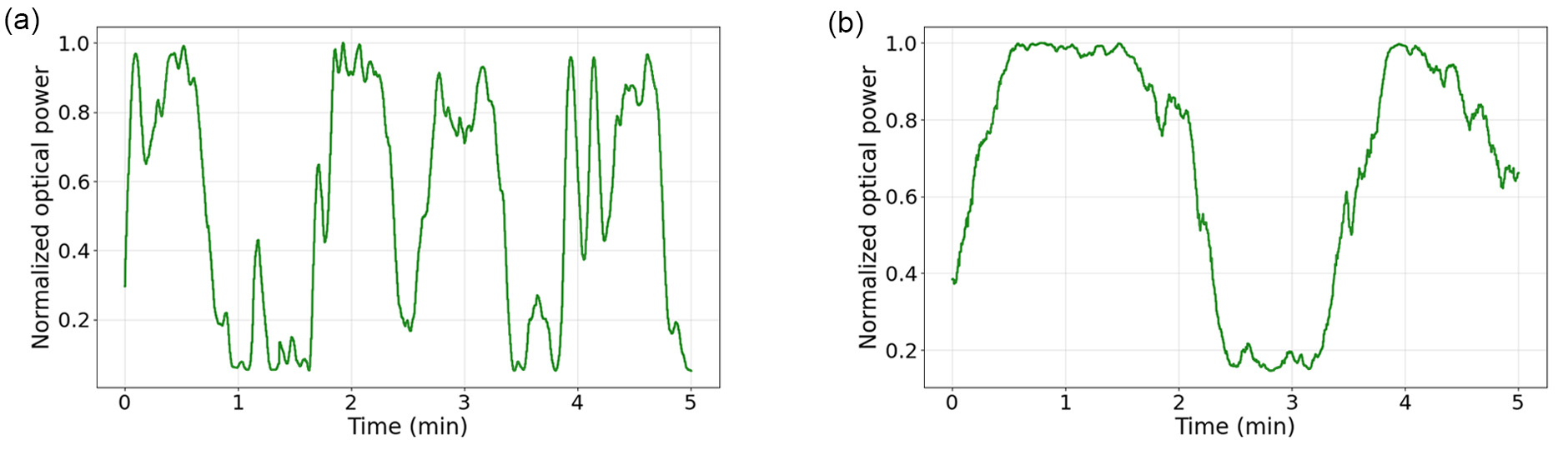}
   \end{center}
   \caption[example] 
   { \label{fig:thermal} 
Phase drifts resulting from environmental fluctuations (a) when the interferometer was not enclosed and  (b) when the cardboard box enclosure was used.}
\end{figure}

Optical fiber interferometers have found several applications as sensors due to their high sensitivity to pressure and temperature variations \cite{dandridge2011}. However, when changes in these parameters are found in the form of environmental fluctuations, e.g., wind currents, mechanical vibrations, and heat emitted by laboratory instruments, the interferometer's performance is affected. This degradation was observed when the interferometer was mounted on an optical bench exposed to the room's environment. Fig. \ref{fig:thermal} shows a time sample of the optical power in one of the outputs when the interferometer was left uncovered and exposed to the open lab environment. While in the linear range of the interferometer $\left( \Delta I\propto\Delta\phi \right)$ drifting has less of an impact, in the nonlinear regime (peaks and troughs), where the nulling interferometer is to operate, phase fluctuations become more impactful for phase stabilization.

In order to minimize background drifting, the interferometer was relocated to the optical bench where the Santa Cruz Extreme AO Lab (SEAL) testbed is located within the same laboratory \cite{jensen-clem2021}. The SEAL testbed is mounted on a heavy granite optical bench and surrounded by an acrylic enclosure, which mitigates contaminating light and reduces hampering environmental perturbations. As a second level of protection, the interferometer setup was covered with a cardboard box. As shown in Fig. \ref{fig:thermal}(b), these modifications improved the phase drifting, increasing the timescale of the relative phase shifts of $\Delta\phi=2\pi$ to approximately 5 minutes.

\subsection{Visibility}
\label{subsec:3-2}
\begin{figure} [h!]
   \begin{center}
   \includegraphics[width=9cm]{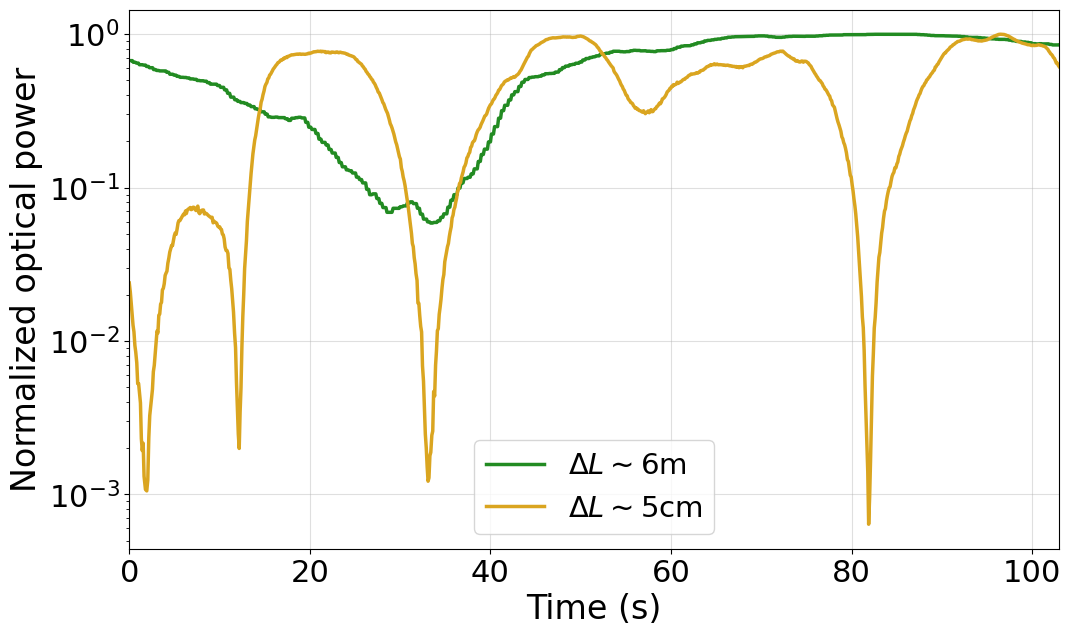}
   \end{center}
   \caption[example] 
   { \label{fig:visib} 
Visibility dependence on path length difference between arms of the interferometer when no active control is being used.}
\end{figure}

The achievable level of null depth is dictated by the contrast of the interference fringes, which in turn will depend on the phase error between the arms of the interferometer. In other words, this refers to how large the relative phase difference between the arms is compared to the coherence length of the source. The visibility, defined as
\begin{equation}
    V = \frac{I_{max} - I_{min}}{I_{max} + I_{min}},
    \label{eq:1}
\end{equation}quantifies this phase error, indicating the path length difference between the arms. During the first stage of experiments with the interferometer, a path difference of $\Delta L\sim$6 m, mostly attributed to the windings in the fiber stretcher, resulted in a limited visibility of $<$90\% as shown in Fig. \ref{fig:visib}. After readjusting the interferometer to the configuration in Fig. \ref{fig:setup}, the path length difference was reduced to $\sim$5 cm, significantly improving the visibility up to levels of 99.8\% (yellow line in Fig. \ref{fig:visib}). Further discussion on the remaining path difference is provided in Sect. \ref{subsec:3-5}. 

\subsection{Optical power matching}
\label{subsec:3-3}
\begin{figure} [h!]
   \begin{center}
   \includegraphics[width=9cm]{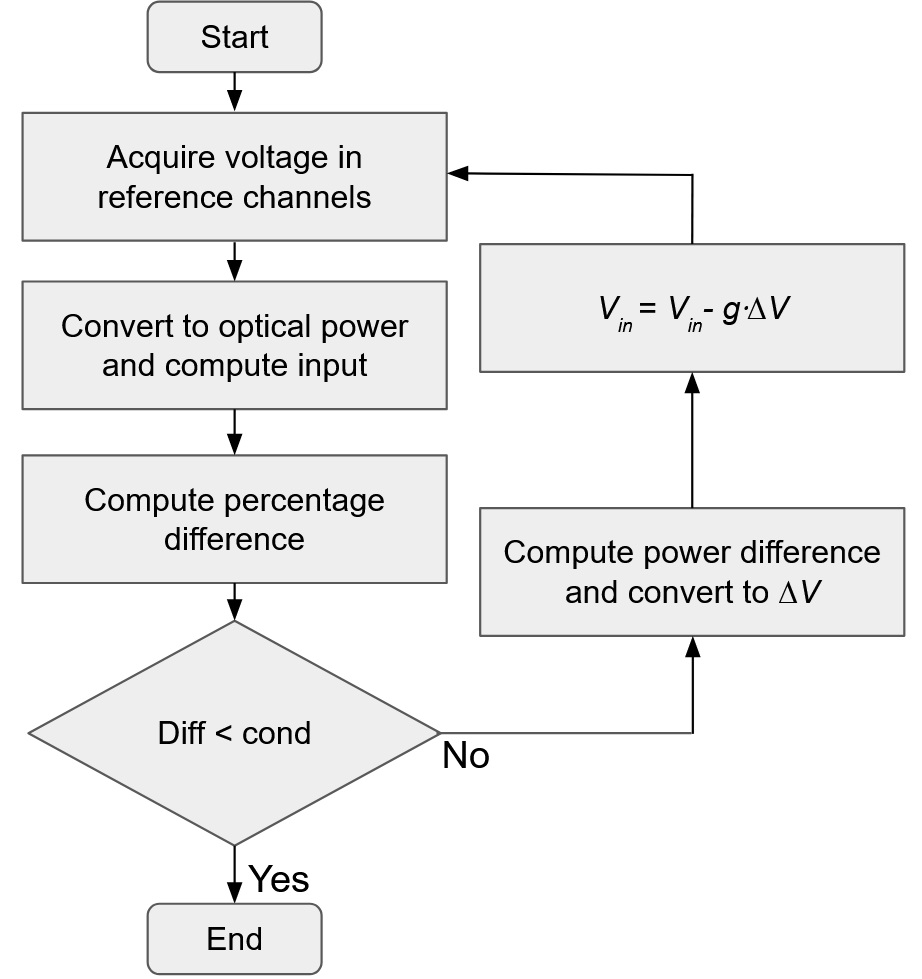}
   \end{center}
   \caption[example] 
   { \label{fig:voa_alg} 
Algorithm employed to match the optical power in the VOA arm to the optical power in the fiber stretcher arm.}
\end{figure}

\begin{figure} [h!]
   \begin{center}
   \includegraphics[width=9cm]{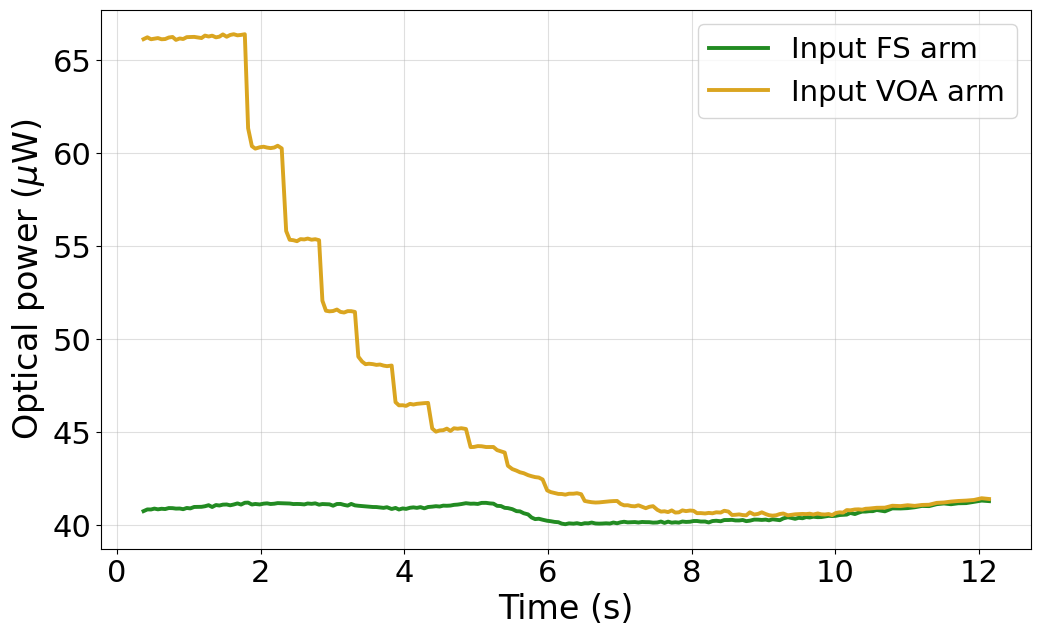}
   \end{center}
   \caption[example] 
   { \label{fig:voaloop} 
Matching the optical power in the VOA arm to the optical power in the fiber stretcher arm.}
\end{figure}

The optical power in the input channels was matched by implementing a simple integrator controller algorithm with the VOA. A block diagram of the algorithm is provided in Fig. \ref{fig:voa_alg}. The photocurrents in the reference photodetectors were monitored and compared until the percentage difference between the input intensities was matched with the VOA, reaching percentage differences $<$0.1\%.  Different values of the integrator's gain were tested, but a thorough analysis of the controller's performance as a function of the gain is still to be done. Also, different speeds of the correction signal, i.e., applied voltage, were tested, which was limited by the power supply's response time, restricting the controller's speed to a few Hz. An example of intensity adjustment is shown in Fig. \ref{fig:voaloop}, for which the loop was closed using a gain of 0.3 at a speed of $\approx$2.5 Hz. 


\subsection{Nulling}
\label{subsec:3-4}
\begin{figure} [h!]
   \begin{center}
   \includegraphics[width=17cm]{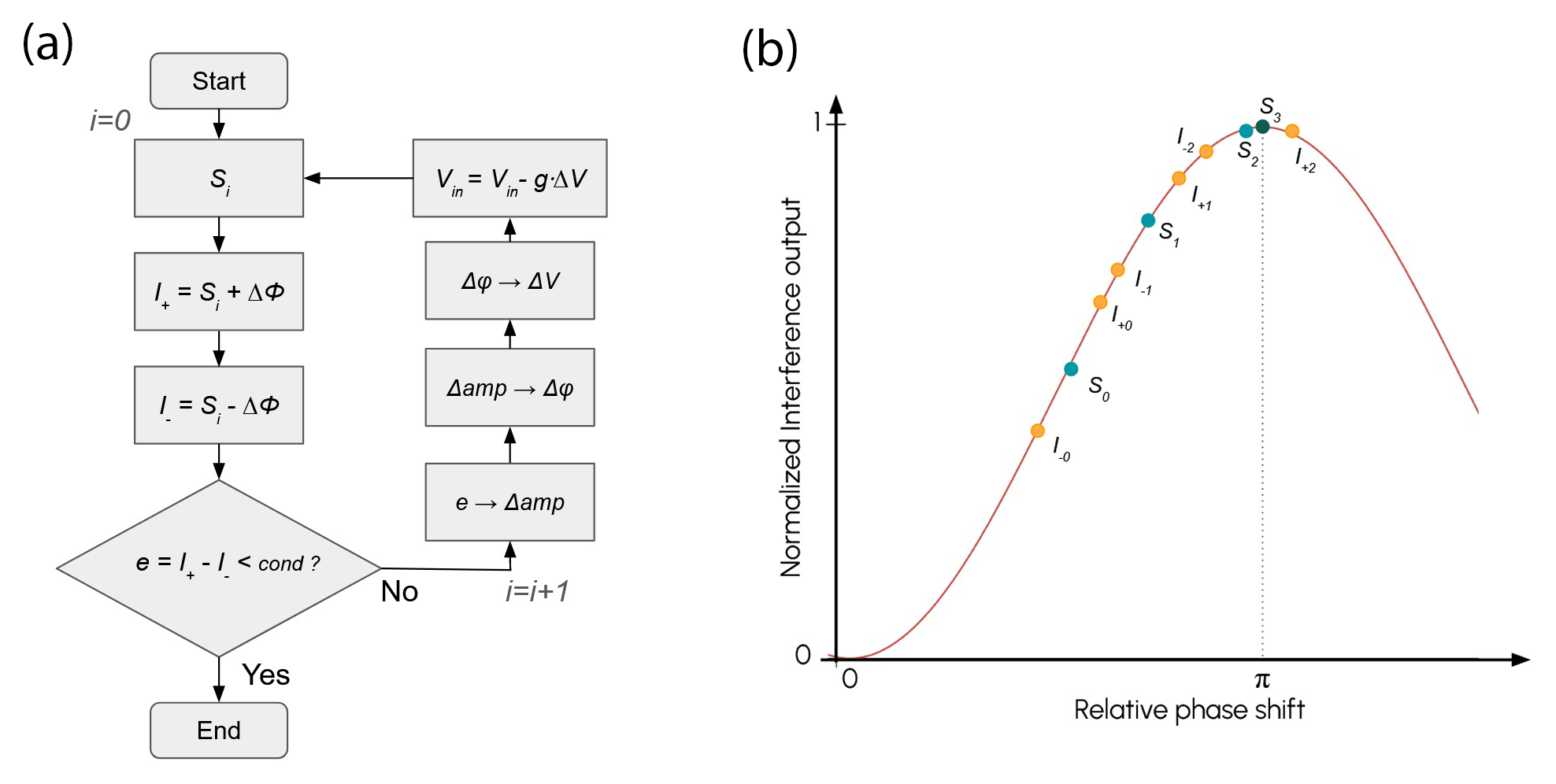}
   \end{center}
   \caption[example] 
   { \label{fig:null_alg} 
(a) Algorithm employed to close the loop with the fiber stretcher to stabilize the nulling. (b) Schematic illustrating how the algorithm approaches a relative phase shift equal to $n\pi$.}
\end{figure}

Having tackled the main factors that could affect the quality of the nulling - visibility, equalizing the amount of light in the inputs, and minimizing background drifting - we moved on to the primary goal of the interferometer: closing the loop with the fiber stretcher to stabilize the nulling. A "dithering" algorithm was implemented to address the degeneracy found around phase values of $n\pi$, using the "bright fringes" channel to track the phase. The "dithering" algorithm was complemented with a simple integrator algorithm to stabilize the phase. Fig. \ref{fig:null_alg}(a) shows a block diagram of the algorithm. In essence, an initial science measurement $\S_{0}$ is taken and is followed by a positive dithering measurement $I_{+,0}$ after applying a positive fixed known phase shift $\Delta\Phi$ (fixed voltage in the fiber stretcher), and a negative dithering measurement $I_{-,0}$ after subtracting $\Delta\Phi$ from the reference $S_{0}$. An error $e=I_{+,0}-I_{-,0}$ is computed and compared to a threshold value. Given the degeneracy around a peak, an error of $e=0$ indicates that the relative phase difference between the arms of the interferometer is a multiple of $\pi$. If the error signal is far from 0, a phase shift $\Delta\phi$ proportional to the amplitude difference between the peak and the current intensity is applied, setting a new science measurement $S_{1}$, and the dithering process is repeated in a loop. The dithering algorithm is illustrated in Fig. \ref{fig:null_alg}(b).

\begin{figure}[h!]
   \begin{center}
   \includegraphics[width=17cm]{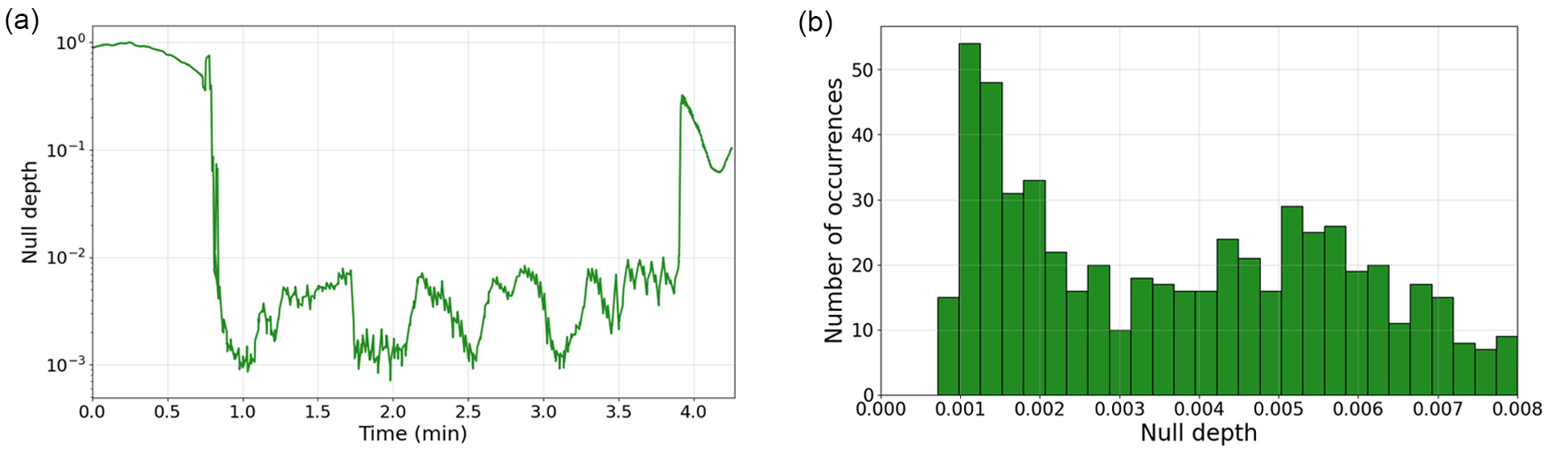}
   \end{center}
   \caption[example] 
   { \label{fig:null1} 
(a) Null depth time series where the null is stabilized for approximately 3 minutes. (b) Histogram of the null depth values measured during the stabilization period.}
\end{figure}

Prior to running the algorithm, the polarization in the VOA arm was adjusted to maximize the interference, and the light intensity in the arms was matched. The null depth \begin{equation}
    N = \frac{I_{min}}{I_{max}},
    \label{eq:2}
\end{equation}
is then computed using for $I_{max}$ the maximum value of optical power before the nulling algorithm is executed. Stabilization of the nulling at levels of $10^{-3}$ was achieved for samples of a few minutes with minimum null depths of $8\times10^{-4}$ reached, as shown in Fig. \ref{fig:null1}. As with the optical power matching procedure, the speed rate of the phase stabilization was limited by the power supply's time response, limiting it to $<$4 Hz. A gain of 0.5 was used in the integrator section for the sample shown here. However, a more detailed examination of the performance as a function of the gain has yet to be done.

\subsection{Performance limitations}
\label{subsec:3-5} 
From our experiments, we identify two performance limitations in our system: the stabilization speed and the minimum achievable null depth. Certainly, the main factor limiting the speed of the stabilization is the power supply's response time. Whereas PyVISA can run communication commands with the instruments in times $<$100 ms, the power supply suffers from a delay in the time it takes to readjust to a new output voltage. Furthermore, communication between the VOA and the power supply is done directly, while for the fiber stretcher, the Model 914-1 piezo-electric controller acts as an intermediary. This results in the degradation of the fiber stretcher compensation speed down to a few Hz. 

Now, to address the performance limitations of the null depth, we first need to know the minimum null depth that we can expect theoretically. In order to do this, we follow the approach presented in Ref.~\citenum{Suchita2017} to evaluate the visibility of a fiber-based Mach-Zehnder interferometer (MZI). As indicated by Eq.\ref{eq:1}, the visibility depends on $I_{max}$ and $I_{min}$, which in turn depend on the coupling ratios of the fiber couplers as
\begin{subequations}
  \begin{equation}
    \label{eq-3a}
      I_{max}=I_{in}\left[\alpha\beta +\left(1-\alpha \right)\left( 1-\beta \right) + 2\sqrt{\alpha\left(1-\alpha \right)\beta \left(1-\beta \right)} \right]
  \end{equation}
  \begin{equation}
    \label{eq-3b}
    I_{min}=I_{in}\left[\alpha\beta +\left(1-\alpha \right)\left( 1-\beta \right) - 2\sqrt{\alpha\left(1-\alpha \right) \beta\left(1-\beta \right)} \right],
  \end{equation}
\end{subequations}
where $\alpha$ is the coupling ratio of the first fiber coupler in the MZI (Fig. \ref{fig:setup}(b)), and $\beta$ is the coupling ratio of the beam combining coupler. Although the first coupler in our setup has a 75/25 splitting ratio, the VOA allows us to adjust the optical power levels to 50/50, hence $\alpha=$0.5. In its turn, the characterization of the beam combining coupler showed a splitting ratio of 52/48 ($\beta=0.52$). Then, by normalizing $I_{in}=$1 and using the previous coupling coefficients, computing $I_{max}$ and $I_{min}$ yields a maximum visibility of $V=$99.9919\% and a minimum null depth of $N=4\times10^{-4}$. Our experimental results show an average null depth of $N\approx3\times10^{-3}$, almost a factor of 3 above the theoretical minimum. The limitation preventing the nuller from achieving its ideal performance can be attributed to the remaining path length difference. Although direct measurement of the optical fiber constituents leads to an estimate of $\Delta l\sim$5 cm, a more accurate number would require a measurement of the bare fiber wound around the fiber stretcher.
\section{CONCLUDING REMARKS AND FUTURE WORK}
\label{sec:4}
We have developed and demonstrated the proof-of-concept stage of an all-fiber-based focal plane nulling interferometer operating at 632 nm. The instrument is composed entirely of off-the-shelf components, including the phase and intensity modulation units, and is controlled and monitored using a PC and the PyVISA package. A fiber stretcher is used to adjust the relative phase shift between the interferometer's arms, yielding stabilization of the null at levels of $10^{-3}$ for periods of a few minutes. The main factors hindering the performance of the nuller are the speed at which the phase compensation can be controlled, limited to only a few Hz due to the response time of the power supply used to control the fiber stretcher, and the remaining path length difference, which prevents from reaching the theoretical minimum null depth.

The instrument's design enables the injection of light into a waveguide at the focal plane and active control of the intensity and phase of the light being transported, an advantage over starlight suppression via a coronagraph alone, which is static. In addition, this design offers the capability to be implemented as a second-stage suppression device downstream of a coronagraph, further increasing the contrast.

The outcomes of this initial demonstration indicate further directions for the concept. Indeed, the first matter to be addressed is overcoming the speed at which the phase is compensated with the fiber stretcher. The next logical step would entail expanding the concept in the laboratory to a multi-wavelength approach and injection from a source simulating the companion. Commercially available SM wavelength division multiplexers could enable the capability to implement fringe tracking at a different wavelength than the one being used for science. Eventually, a major milestone implies on-sky testing of the instrument. A potential route for this is through the APALO (Astrophotonics Advancement at Lick Observatory), which builds on our team's work with an earlier platform called the Parallel Lantern Injection Unit (PLIU) on the Shane 3-m Telescope at Lick Observatory \cite{DeMartino2022}. The PLIU platform was designed to inject AO-corrected light simultaneously into two photonic lanterns. Coupling of light coming from the telescope into a photonic lantern downstream of the ShaneAO system has been demonstrated (see DeMartino et al. 2024 in these proceedings), providing avenues to feed the nuller and future astrophotonics instrumentation.

\acknowledgments 
   
We thank the University of California Observatories for funding this research. J. Diaz thanks Vincent Chambouleyron for the suggestions and discussions that contributed to making this project possible, and the Cota-Robles Fellowship for all the support provided.
\bibliography{report} 
\bibliographystyle{spiebib} 

\end{document}